\documentclass[aps,pra,reprint, amsmath, amssymb,superscriptaddress]{revtex4-1}

\usepackage{bm}
\usepackage[retainorgcmds]{IEEEtrantools}
\usepackage{graphicx}
\usepackage{mathrsfs}
\usepackage{amsmath}
\usepackage{amssymb}
\usepackage{color}
\usepackage{amsfonts}
\usepackage{amsthm}
\usepackage{enumerate}
\usepackage{dsfont}
\usepackage{nicefrac}

\DeclareMathOperator{\tr}{tr}

\newlength{\eqboxstorage}

\begin{document}

\title{Thermodynamic analysis of quantum error correcting engines}
\date{\today}
\author{Gabriel T. Landi}
\email{gtlandi@if.usp.br}
\affiliation{Instituto de F\'isica da Universidade de S\~ao Paulo,  05314-970 S\~ao Paulo, Brazil.}
\author{Andr\'e L. Fonseca de Oliveira}
\affiliation{Facultad de Ingenier\'ia, Universidad ORT Uruguay, Uruguay}
\author{Efrain Buksman}
\affiliation{Facultad de Ingenier\'ia, Universidad ORT Uruguay, Uruguay}

\begin{abstract}
Quantum error correcting codes can be cast in a way which is strikingly similar to a quantum heat engine undergoing an Otto cycle. 
In this paper we strengthen this connection further by carrying out a complete assessment of the thermodynamic properties of 4-strokes operator-based error correcting codes. 
This includes an expression for the entropy production in the cycle which, as we show, contains clear contributions stemming from the different sources of irreversibility. 
To illustrate our results, we study a classical 3-qubit error correcting code, well suited for incoherent states, and the 9-qubit Shor code 
capable of handling fully quantum states.  
We show that the work cost associated with the correction gate is directly associated with the heat introduced by the error. 
Moreover, the  work cost associated with encoding/decoding  quantum information is always positive, a fact which is related to the intrinsic irreversibility introduced by the noise. 
Finally, we find that correcting the coherent (and thus genuinely quantum) part of a quantum state introduces substantial modifications related to the Hadamard gates required to encode and decode coherences.

\end{abstract}
\maketitle{}

%
%
%
%
\section{Introduction}

\begin{figure}[!h]
\centering
\includegraphics[width=0.45\textwidth]{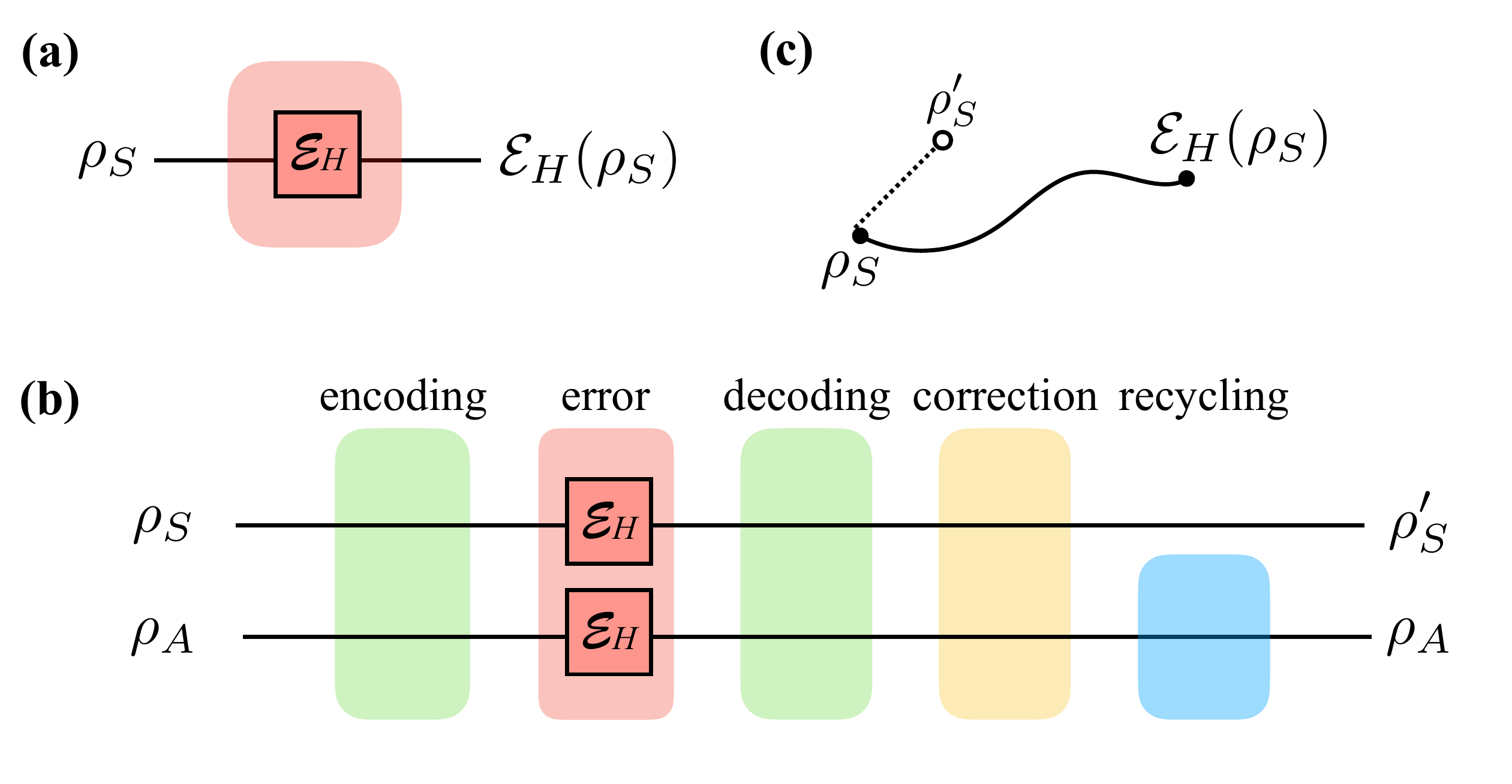}
\caption{\label{fig:drawing}
Typical error correcting scenario. 
(a) A state $\rho_S$ is susceptible to error, described by a quantum channel $\mathcal{E}_H(\rho_S)$. 
(b) In order to protect it, $\rho_S$ is first encoded into ancillas $A$. 
After both undergo individual errors $\mathcal{E}_H$, the state of the system is decoded from $S+A$ and  a set of correction measures are applied, leading to a final state $\rho_S'$  for the system. 
This procedure makes the ancillas dirty, which must then be recycled if they are to be used again. 
(c) The code is considered successful (at the ensemble level) whenever it mitigates the role of the noise, which means $D(\rho_S',\rho_S) < D(\mathcal{E}_H(\rho_S), \rho_S)$, where $D$ is any distance measure.
This, of course, will be the case only if the effect of $\mathcal{E}_H$ is sufficiently small.
}
\end{figure}

Quantum error correcting codes (QECCs)  protect qubits from detrimental noise by redundantly storing quantum states in multiple parties \cite{Lidar2013}. 
The basic idea is illustrated in Fig.~\ref{fig:drawing}. 
An environment induces noise in a system $S$, which is modeled as a quantum channel  $\mathcal{E}_H$, as in Fig.~\ref{fig:drawing}(a). 
In order to protect it, the state of the system is encoded into a larger Hilbert space by introducing additional ancillas. 
Both system and ancillas are now susceptible to the noise process. 
But by applying appropriate correction measures, one may mitigate this noise at the expanse of making the final state of the ancillas more mixed 
 (Fig.~\ref{fig:drawing}(b)).

The connection between QECCs and thermodynamics has been discussed for quite some time in connection with Landauer's erasure and Maxwell's Demon \cite{Barbara1998,Vedral2000a}. 
However, an interesting connection which, to our knowledge  has never been explored, is that with quantum heat engines (QHE) \cite{Kosloff2014,Mitchison2019}. 
This becomes  accurate in the case of operator error correction  \cite{Kribs2005,Kribs2006,Clemens2006,Tomita2011}, where no syndrome measurements are required. 
The diagram in Fig.~\ref{fig:drawing} is then seen to be entirely analogous to a quantum heat engine undergoing an Otto cycle: 
The ``working fluid''   is composed of both  system and ancillas.
The encoding, decoding and correction steps are the unitary strokes, involving the expenditure of work without any heat flow. 
The noise term represents the action of the hot bath. 
And finally, the recycling step where the states of the ancillas are reset, represents the cold bath. 

In view of this striking similarity, one is naturally led to ask how far can this connection be pushed. 
Of course, in the end, the goal of a QECC is entirely different from that of a QHE. 
Efficiency, for instance, has nothing to do with work extraction, but with the ability of the code to correct the error. 
Notwithstanding these fundamental differences, an analysis of a QECC from a thermodynamic perspective is still  illuminating, as it allows one to address the roles of heat and work in the error correcting process. 
Particularly interesting is the question of what is the work cost for encoding and decoding quantum information, 
as compared to the cost for applying an error correction. 
For instance, is it possible to successfully apply an error correction and still extract useful work from the machine? 
Or does the success of the QECC necessarily involve the expenditure of work by an external agent?

With these motivations in mind, in this paper we put forth a complete thermodynamic characterization of QECCs implemented using the operator error correction scheme \cite{Kribs2005,Kribs2006}. 
We begin by considering the general thermodynamic properties, including a reformulation of the first and second laws for the specific QECC scenario. 
Next we apply these results to two representative examples. 
The first is a 3-qubit classical error correction, capable of correcting incoherent states. 
The second is the fully quantum 9-qubit Shor code, which can simultaneously handle both incoherent as well as coherent states. 


%
%
%
%
\section{Formal Framework}

In this section we provide a general characterization of the thermodynamic properties of the QECC in Fig.~\ref{fig:drawing}.
We begin by describing the basic strokes of the cycle and then move on to characterize it in terms of the first and second laws of thermodynamics. 

\subsection{Description of the cycle}

We assume the main system $S$ is a qubit with computational basis $|0\rangle, |1\rangle$ and Hamiltonian $H_S = \frac{\Omega}{2} (1- \sigma_z^S)$ (so that $|0\rangle$ is the ground-state). 
The code involves coupling the system with a set of ancillas, which we shall henceforth assume to be identical, with Hamiltonian $H_{A_i} = \frac{\omega}{2} (1- \sigma_z^{A_i})$. 
The ancillas are always prepared in the ground state $|0\rangle$, so that the global initial state of $N$ ancillas is 
$\rho_A = |0\rangle\langle 0 |^{\otimes N}$.
Below, when convenient, we shall assume for simplicity that $\omega = \Omega$. 

In this paper we will consider 4-stroke codes, each of which we now explain in detail. 
The first stroke is the \emph{encoding stroke}, where the system density matrix $\rho_S$ is encoded in the ancillas by means of a unitary $U_e$,
\begin{equation}\label{stroke_1}
\rho_{SA}^{(1)} = U_e \big( \rho_S \otimes \rho_A) U_e^\dagger. 
\end{equation}
The second stroke is the \emph{error (noise) stroke}, where both $S$ and $A$ are subject to local noise channels.
In order to highlight the correction with thermodynamics, we consider the noise generated by the generalized amplitude damping (GAD) channel, 
\begin{equation}\label{GAD}
\mathcal{E}_H(\rho) = \sum\limits_{k=1}^4 M_k \rho M_k^\dagger, 
\end{equation}
where 
\begin{IEEEeqnarray}{LL}
M_1 = \sqrt{1-f} \begin{pmatrix} 1 & 0 \\ 0 & \sqrt{1-\gamma}\end{pmatrix},
&
M_2 = \sqrt{1-f} \begin{pmatrix} 0 & \sqrt{\gamma} \\ 0 & 0\end{pmatrix},
\nonumber \\[0.2cm]
M_3 = \sqrt{f} \begin{pmatrix} \sqrt{1-\gamma} & 0 \\ 0 & 1\end{pmatrix},
&
M_4 = \sqrt{f} \begin{pmatrix} 0 & 0 \\ \sqrt{\gamma} & 0\end{pmatrix}.
\end{IEEEeqnarray}
Here $\gamma \in [0,1]$ is the coupling strength and $f$ is excited state probability (Fermi-Dirac distribution). 
If $f = 0$ the map will target the ground-state $|0\rangle$.  
Since $S$ and $A$ have different frequencies, we will use the notation $f_x = (e^{\beta_H x} + 1)^{-1}$, with $x = \Omega, \omega$ and $\beta_H$ being the temperature of the hot bath.
Error correction is mostly successful when the noise strength $\gamma \ll 1$, which we shall assume  throughout this paper. 
Moreover, following customary treatments of error correction, all results for specific codes will be presented in terms of a power series, only to leading order in $\gamma$. 
The state after the second stroke will be
\begin{equation}\label{stroke_2}
\rho_{SA}^{(2)} = \mathcal{E}_H^S \otimes \mathcal{E}_{H}^{\otimes N} (\rho_{SA}^{(1)}).
\end{equation}

The third stroke is the \emph{decoding/correction} operation. 
This will again be described by a unitary $U_{dc}$, which in general cannot be split as the product of two unitaries for decoding and correction. 
The state after the third stroke will be 
\begin{equation}\label{stroke_3}
\rho_{SE}^{(3)} = U_{dc} \rho_{SA}^{(2)} U_{dc}^\dagger.
\end{equation}
Finally, the fourth stroke is the \emph{recycling} stroke, where the ancillas interact with a cold bath and nothing is done to the system. 
This stroke can also be viewed  as the action of a GAD~(\ref{GAD}), but with $\gamma = 1$ and $f = 0$. 
However, this is not necessary since its effect is simply to reset the state of the ancillas. 
Hence, after the fifth stroke the global state will be
\begin{equation}\label{stroke_4}
\rho_{SA}^{(4)} = \rho_S^{(3)} \otimes \rho_A, 
\end{equation}
where $\rho_{S}^{(3)} = \tr_A \rho_{SA}^{(3)}$ is the state of $S$ after the fourth stroke and $\rho_A$ was the initial state of the ancillas. 

\subsection{Error correcting efficiency}

For conciseness, we shall denote by $\rho_S' = \rho_S^{(3)} = \rho_{S}^{(4)}$ as the final state of the system after one cycle. 
Thus, from a global perspective the input state of the engine is $\rho_S \otimes \rho_A$ and the output state is $\rho_S' \otimes \rho_A$. 
We therefore see that, in general, the engine's operation is not cyclic (i.e., it has not reached a limit cycle). 
This, actually, is precisely what quantifies the efficiency of the error correcting code, as the goal of the engine is to have $\rho_S'$ as close as possible to $\rho_S$. 

Motivated by this one can define the efficiency of the QECC as follows. Let $D(\rho,\sigma)$ denote any proper distance measure between quantum states. 
To address the success of a QECC, one must compare $\rho_S'$ with the state $\mathcal{E}_H(\rho_S)$ which one would obtain if only the error map $\mathcal{E}_H$ were to be applied to the state. 
A QECC can be declared successful (at the ensemble level) if
\begin{equation}
D(\rho_S', \rho_S) < D(\mathcal{E}_H(\rho_S), \rho_S), 
\end{equation}
since this implies that the effect of the noise was at least partially mitigated by the code. 
Hence, a proper measure of the efficiency of a QECC could be, for instance, 
\begin{equation}\label{efficiency}
\eta_\text{\tiny QECC} = 1 - \frac{D(\rho_S',\rho_S)}{D(\mathcal{E}_H(\rho_S),\rho_S)}.
\end{equation}
This quantity is 1 when the correction is perfect, zero when the correction has no effect and negative when the code actually makes things worse. 
It resembles the thermodynamic efficiency, but is purely information-theoretic. 
Below we will not need this specific form of the QECC efficiency in order to construct the cycle's thermodynamic. 
We have presented it here simply to emphasize that the QECC efficiency is, in general, not at all related to any thermodynamic efficiency.

\subsection{First law of thermodynamics}

The operations described by the four strokes in Eqs.~(\ref{stroke_1})-(\ref{stroke_2}) are essentially implementing an Otto cycle. 
Strokes 1 and 3 are unitary, involving the possible expenditure of work, but without any exchange of heat. 
Similarly, strokes 2 and 4 are purely dissipative, involving only the exchange of heat and no work. 
The expressions for the heat and work in each stroke are thus easily calculated as the changes in energy in each stroke, 
$W_e = \Delta H_{10}$, $Q_H = \Delta H_{21}$, $W_{dc} = \Delta H_{32}$ and $Q_C = \Delta H_{43}$, where $H = H_S + H_A$ is the total Hamiltonian and $\Delta H_{i,i-1} = \tr\big\{ H(\rho_{SA}^{(i)} - \rho_{SA}^{(i-1)})\big\}$ is the total change in energy of each stroke. 

The decoding/correction stroke~(\ref{stroke_3}) is described by a unitary $U_{dc}$ which can be decomposed as a riffling   $U_{dc} = U_d^{(1)} U_c^{(1)} U_d^{(2)} U_c^{(2)}\ldots$, where $U_d^{(i)}$ and $U_c^{(i)}$ represent decoding and correcting steps respectively. 
Based on this, the work $W_{dc}$ can also be split as $W_{dc} = W_d + W_c$, giving the individual contributions from each part of the code. 

The ancillas are reset after each stroke, but the system is not. Hence, as a consequence, the first law of thermodynamics reads
\begin{equation}\label{first_law}
\Delta U_S = W_e + Q_H + W_{dc} + Q_C,
\end{equation}
where $\Delta U_S = \tr\big\{ H_S (\rho_S' - \rho_S)\big\}$ is the change in energy of the system only.
Since the total Hamiltonian is split as $H = H_S + H_A$, the same may also be done for all heat and work contributions. 
Thus, we may also write the first law as
\begin{equation}\label{first_law_2} 
\Delta U_S = \big( W_e^S + Q_H^S + W_{dc}^S\big) + \big( W_e^A + Q_H^A + W_{dc}^A + Q_C^A\big), 
\end{equation}
where we used the fact that the heat to the cold bath only has an ancilla part. 
But since the operations are all local and since the state of the ancillas are reset, it follows that the last term must be identically zero. 
Hence, the first law can be written solely as a system-related quantity:
\begin{equation}
\Delta U_S = W_e^S + Q_H^S + W_{dc}^S.
\end{equation}

\subsection{Second law  for the noise stroke}

One can also write down the second law of thermodynamics for the QECC. 
The heating stroke 2 involves a standard finite-temperature amplitude damping, for which the expression for the entropy production is very well established \cite{Esposito2010a,Reeb2014,Brandao2015,Manzano2017a,Strasberg2016} and reads
\begin{equation}\label{Sigma}
\Sigma_H = \Delta S_{21} - \beta_H Q_H \geq 0, 
\end{equation}
where $\Delta S_{21} = S(\rho_{SA}^{(2)}) - S(\rho_{SA}^{(1)})$ is the change in von Neumann entropy ($S(\rho) = -\tr(\rho \ln \rho)$) in stroke 2.
The positivity of $\Sigma_H$ is a direct consequence of the data processing inequality \cite{Breuer2003}.

This  expression can be manipulated so as to better highlight the physical origins of the irreversibility associated with the QECC cycle. 
Since stroke 3 is unitary, it follows that $S(\rho_{SA}^{(2)}) = S(\rho_{SA}^{(3)})$. 
Moreover, we can write 
\[
S(\rho_{SA}^{(3)}) = S(\rho_S') + S(\rho_A^{(3)}) - \mathcal{I}^{(3)}(S:A),
\]
where $\mathcal{I}^{(3)}(S:A)$ is the mutual information between system and ancilla in the state $\rho_{SA}^{(3)}$. 
Similarly, since the first stroke is unitary, $S(\rho_{SA}^{(1)}) = S(\rho_S) + S(\rho_A) = S(\rho_S)$, as the ancillas are taken to be in a pure state. 
Whence, the entropy production~(\ref{Sigma}) can be written as 
\begin{equation}\label{Sigma_2}
\Sigma_H = \Delta S_S + S(\rho_A^{(3)}) - \mathcal{I}^{(3)}(S:A) - \beta_H Q_H \geq 0.
\end{equation}
This is an important result. 
The first term is the total change in entropy of the system, $\Delta S_S = S(\rho_S') - S(\rho_S)$. 
It is precisely one of the goals of the QECC  to minimize $\Delta S_S$.  
The second term in Eq.~(\ref{Sigma_2}) is the entropy increase in the ancillas. 
As a byproduct of the QECC, the ancillas become dirty, which is precisely quantified by this term. 
Hence, $S(\rho_A^{(3)})$ will be exactly the amount of entropy that has to be cleaned up in the last recycling stroke. 

The third term in Eq.~(\ref{Sigma_2}) is the \emph{residual} mutual information that still remains between system and ancilla after the decoding/correction stroke. 
In the limit of perfect correction, the system would return to $\rho_S$, so that $\mathcal{I}^{(3)}(S\!:\! A) = 0$. 
Hence, $\mathcal{I}^{(3)}(S\!:\!A)$ represents the shared information that remained in the state $\rho_{SA}^{(3)}$ which the correcting scheme was unable to remove. 
This mutual information appears with a negative sign, hence contributing to make the process more reversible. 
The reason for this lies in the fact that before the recycling stroke $\mathcal{I}^{(3)}(S:A)$ is still in principle accessible. 
As we shall see below, once one includes the recycling stroke, however, these correlations are irretrievably lost. 

Finally, the last term in Eq.~(\ref{Sigma_2}) is the heat flow to the hot bath. 
Since $Q_H = Q_H^S + Q_H^A$, we may also write~(\ref{Sigma_2}) more symmetrically as 
\begin{equation}\label{Sigma_3}
\Sigma_H = (\Delta S_S - \beta_H Q_H^S) + (S(\rho_A^{(3)}) - \beta_H Q_H^A) - \mathcal{I}^{(3)}(S\!: \!A),
\end{equation}
which is clearly split into two local contributions, plus a genuinely non-local term. 

\subsection{Second law for the recycling stroke}

One can similarly write down the second law for the interaction with the cold bath. 
In this case, however, an equation of the form~(\ref{Sigma}) would give diverging results, as $\beta_C = \infty$. 
This pathological behavior of the entropy production in the limit of zero temperature is a known issue, which was discussed for instance in Refs~\cite{Santos2018,Santos2017b,Santos2018a}. 
To circumvent, one must provide additional details on  the environment interaction generating the map. 
We therefore assume that each ancilla $A_i$ is coupled to a corresponding environment $E_i$ (not necessarily qubits) prepared in a pure state $|0\rangle_{E_i}$, while the system $S$ is not coupled to anything. 
We assume in this stroke that the ancillas are fully reset back to $|0\rangle_{A_i}$, which means that each  $A_iE_i$ interaction must have the form of a full SWAP. 
With this proviso, the recycling stroke may be written as the map composition
\begin{equation}
\rho_{SA}^{(4)} = \mathcal{E}_C^{(A_1)} \otimes \ldots \otimes \mathcal{E}_C^{(A_N)} (\rho_{SA}^{(3)}),
\end{equation}
where
\begin{equation}
\mathcal{E}_C^{(A_i)}(\rho) = \tr_{E_i} \bigg\{ U_{A_i,E_i}^\text{SWAP} \bigg( \rho \otimes |0\rangle\langle 0 |_{E_i} \bigg) (U_{A_i,E_i}^\text{SWAP})^\dagger \bigg\}, 
\end{equation}
is the Stinespring dilation for a map acting only on ancilla $A_i$. 

With this specific representation for the recycling stroke, we can now propose a formula for the entropy production. 
Namely, based on Refs.~\cite{Strasberg2016,Esposito2010a,Manzano2017a}, we define the entropy production  as being only the mutual information between $SA$ and the cold environment $E$.  
\begin{equation}\label{Sigma_C}
\Sigma_C = \mathcal{I}(SA:E) = S(\rho_{SA}^{(4)}) + S(\rho_E') - S(\rho_{SAE}'),
\end{equation}
where $\rho_{SAE}'$ denotes the global state of system, ancillas and cold environment after the map, with $\rho_E'$ being the corresponding reduced density matrix. 
Within the context of  dilated unitary maps, entropy production is often defined with an additional term, proportional to the relative entropy between the initial and final states of the environment \cite{Esposito2010a}. 
In fact, quite recently this extra term was shown to be extremely important in a large variety of models \cite{Ptaszynski2019b}.
However, in the case of zero temperature, it gives a diverging result since the initial state of the environment is pure. 
The expression~(\ref{Sigma_C}), which is discussed also in \cite{Strasberg2016,Manzano2017a}, is a choice that does not suffer from this pathology. 

Since the global $SAE$ dynamics is unitary, it follows that $S(\rho_{SAE}') = S(\rho_{SA}^{(3)}) + S(\rho_E) = S(\rho_{SA}^{(3)})$. 
Moreover, we are assuming full thermalization so that $S(\rho_{SA}^{(4)}) = S(\rho_{S}^{(3)})$. 
And, finally, again because of the assumption of full thermalization, $S(\rho_E') = S(\rho_A^{(3)})$. 
Whence, we conclude that  Eq.~(\ref{Sigma_C}) may also be written as 
\begin{equation}\label{Sigma_C_2}
\Sigma_C = S(\rho_S^{(3)}) + S(\rho_{A}^{(3)}) - S(\rho_{SA}^{(3)}) = \mathcal{I}^{(3)} (S\! : \! A). 
\end{equation}
This result shows that the entropy production in the cold stroke is nothing but the residual mutual information that was developed between system and ancillas in the previous strokes, and which is lost due to the action of the cold bath. 
This is the same residual mutual information appearing in Eq.~(\ref{Sigma_3}).

Combining Eqs.~(\ref{Sigma_3}) and (\ref{Sigma_C_2}) then finally leads to a formula for the total entropy production in the QECC engine:
\begin{IEEEeqnarray}{rCl}
\Sigma &=& \Sigma_H + \Sigma_C \nonumber\\[0.2cm]
&=& \Delta S_S - \beta_H Q_H^S + S(\rho_A^{(3)})  - \beta_H Q_H^A.
\end{IEEEeqnarray}
Whence, the total entropy production is found to contain only \emph{local} contributions, referring to the changes taking place in system and ancilla.

\begin{figure}
\centering
\includegraphics[width=0.45\textwidth]{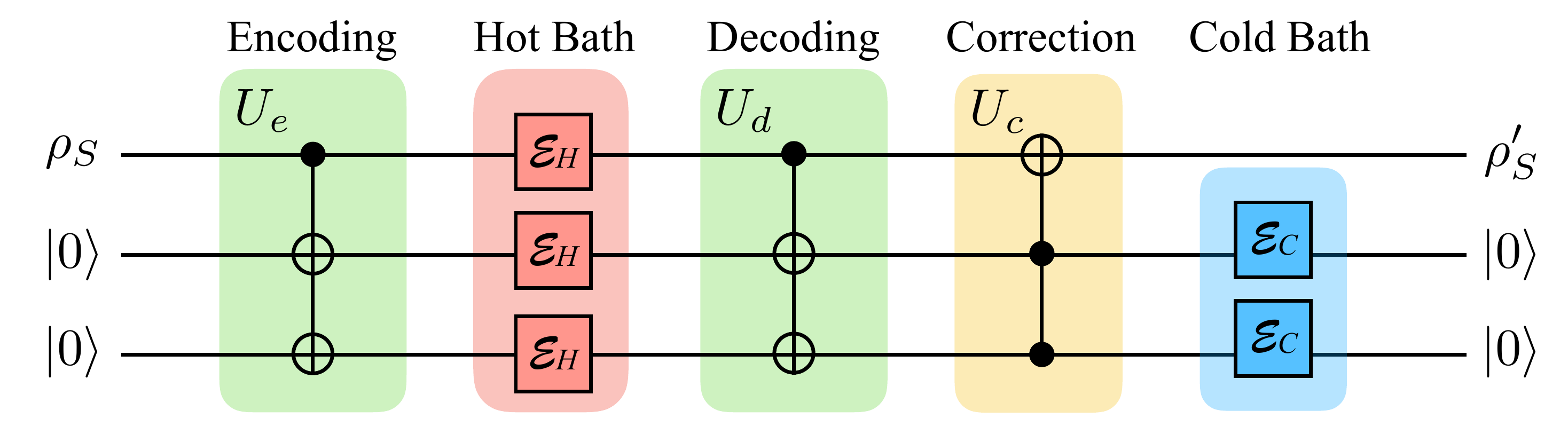}
\caption{\label{fig:classical}
The classical error correcting algorithm capable of correcting for diagonal states of the system. 
}
\end{figure}

%
%
%
%
\section{\label{sec:classical}Classical error correcting engine}

In the remainder of the paper, we apply our results to two specific QECCs. 
To start, we consider the particularly illuminating case of \emph{classical} error correction. 
That is, we first consider the protection of diagonal states (in the computational basis) of the form 
\begin{equation}\label{classical_state}
\rho_S = (1-p) |0\rangle\langle 0 | + p |1\rangle \langle 1 |, \qquad p \in [0,1].
\end{equation}
This state can be regarded as classical as far as the amplitude damping channel is concerned since, in the sense of einselection \cite{Zurek1981,Zurek2003b}, the amplitude damping chooses the computational basis as a preferred basis. 

The effects of the amplitude damping on the state~(\ref{classical_state}) can be corrected by the 3-qubit majority voting scheme shown in Fig.~\ref{fig:classical}(a).
The encoding unitary $U_e$ is composed of a double CNOT, 
\begin{equation}
U_e = |0\rangle\langle 0 |_S \otimes I_{A_1} \otimes I_{A_2}+|1\rangle\langle 1 |_S \otimes X_{A_1} \otimes X_{A_2},
\end{equation}
where $X=\sigma_x$ is the Pauli operator. 
Moreover, the decoding/correction unitary $U_{dc}$ in this case is factored into a product of two terms, $U_{dc} = U_d U_c$, with $U_d = U_e$ and $U_c$ being a Toffoli gate.

All strokes can be computed using standard symbolic algebra. 
We begin by considering the fidelity between the final and initial states of the system, with and without the QECC.
In this case we assume for simplicity that $\omega = \Omega$, so we can set  $f_\Omega = f_\omega \equiv f$. 
If no QECC is applied we find, to leading order in the noise strength $\gamma$, 
\begin{equation}
F(\mathcal{E}_H(\rho_S), \rho_S) \simeq 1 - \frac{\gamma ^2 }{4 (1-p) p}(f-p)^2.
\end{equation}
As expected, the fidelity is unity if $p = f$, in which case the system already starts with the same population as the environment. 
Conversely, if the QECC  is applied to protect the system one finds that 
\begin{equation}
F(\rho_S', \rho_S) \simeq 1 - \frac{9\gamma^4}{4(1-p)p} \bigg[ 
p(1-2f)-f^2(1-2p)
\bigg]^2,
\end{equation}
We see that the leading term in the fidelity when the QECC is applied now becomes  $\sim\gamma^4$, as compared to $\gamma^2$ without the QECC.
This neatly shows how error correction behaves at the ensemble level. 

\begin{figure}[!ht]
\centering
\includegraphics[width=0.5\textwidth]{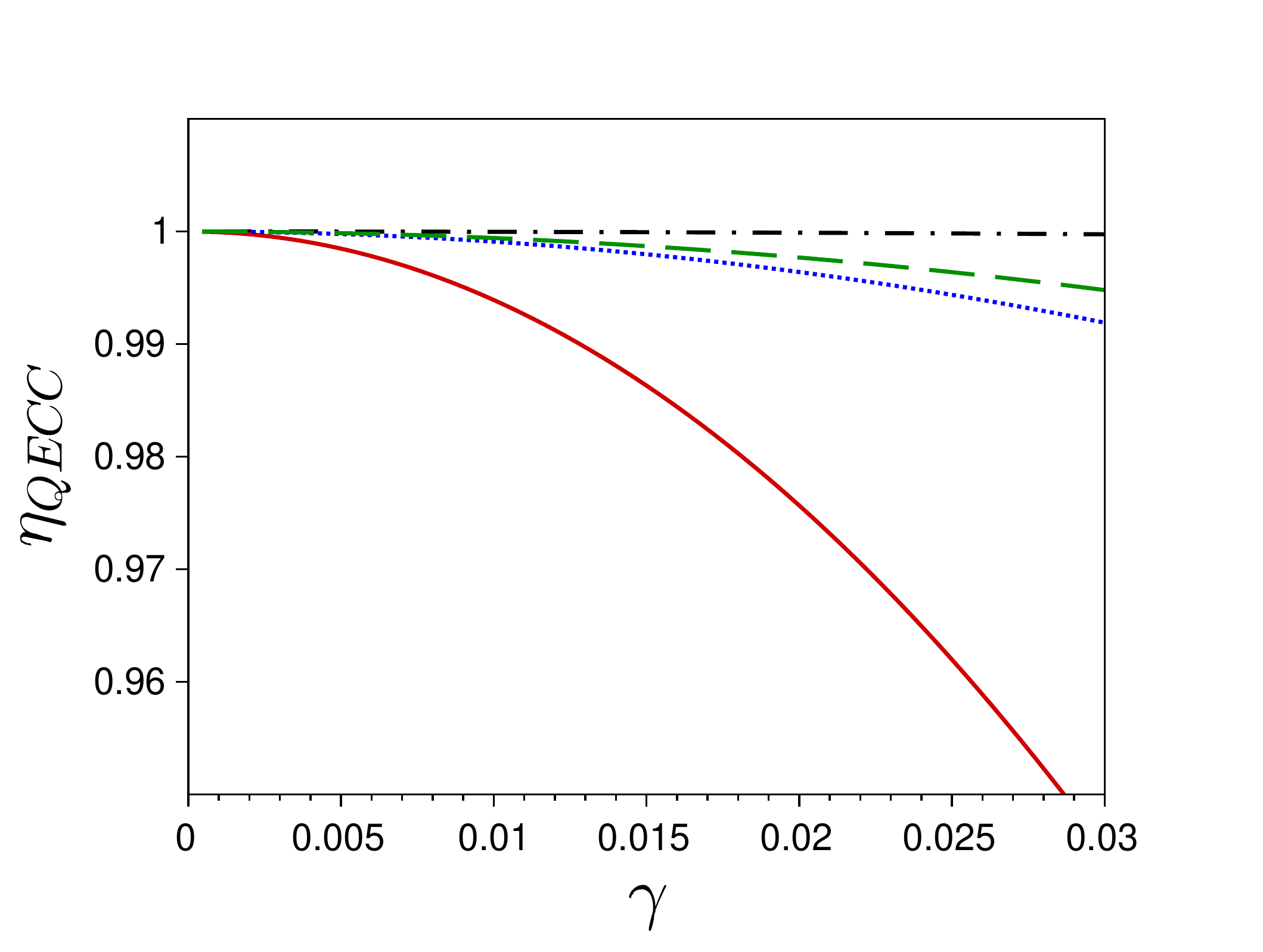}
\caption{\label{fig:effic_3qbs_1}
Efficiency for states with $p=0.01$ (dash-dot), $p=0.99$ (dash), $p=0.5$ (dot) and $p=0.25$ (solid). For all cases $f=0.2$. 
}
\end{figure}

Let us now compute the efficiency defined in Eq.~(\ref{efficiency}). 
As a proper distance measure we use the Bures distance squared, 
\begin{equation}
D^2(\rho, \sigma) = 2\bigg(1- \sqrt{F(\rho,\sigma)}\bigg).
\end{equation}
The efficiency~(\ref{efficiency}), to leading order in $\gamma$, then becomes
\begin{equation}
\label{eq:effic_3qbs}
\eta_\text{\tiny QECC} \simeq 1 - \frac{ 9 \gamma^2}{(f-p)^2} \bigg[ p(1-2f)-f^2(1-2p) \bigg]^2. 
\end{equation}
We see that error correction becomes problematic when $p \to f$, as in this case the effect of the channel becomes trivial, so that there is no error to correct. Fig.~ \ref{fig:effic_3qbs_1} shows the behavior of the efficiency using Eq.~(\ref{eq:effic_3qbs}).

Next we present the heat and work in each step, which we divide into contributions from the system and from the ancillas. 
The contributions referring to the system, again, to leading order in $\gamma$, are 
\begin{IEEEeqnarray}{rCl}
\label{classical_We_S}
W_e^S &=& 0 ,		\\[0.2cm]
\label{classical_QH_S}
Q_H^S &=& \gamma \Omega (f_\Omega - p) , 	\\[0.2cm]
\label{classical_Wd_S}
W_d^S &=& 0,\\[0.2cm]
\label{classical_WcS}
W_c^S &\simeq& - \gamma \Omega(f_\Omega-p), 		
\end{IEEEeqnarray}
whereas the contributions from the ancillas are 
\begin{IEEEeqnarray}{rCl}
\label{classical_We_A}
W_e^A &=& 2 p \omega, 	\\[0.2cm]
\label{classical_QH_A}
Q_H^A &=& 2 \gamma \omega (f_\omega - p), 	\\[0.2cm]
\label{classical_Wd_A}
W_d^A &\simeq& - 2 p \omega + 2 \gamma \omega \bigg[ f_\Omega +p (3-  2 f_\omega -  2 f_\Omega)\bigg],		\\[0.2cm]
\label{classical_Wc_A}
W_c^A &=& 0, 		\\[0.2cm]
\label{classical_QC_A}
Q_C^A &\simeq& -2 \gamma \omega \bigg[ f_\Omega +p (3-  2 f_\omega -  2 f_\Omega)\bigg] - 2 \gamma \omega(f_\omega - p).
\IEEEeqnarraynumspace
\end{IEEEeqnarray}

The physics behind each term is quite interesting. 
First, the work $W_e$ of the encoding stroke is only associated with thecost of putting the two ancillas in the excited state with probability $p$. 
Next, the heat that flows to the hot bath is proportional to the population mismatch $f_\Omega - p$ and $f_\omega - p$. 
It may thus have any sign depending on the initial value of $p$. 
Hence, it is very well possible for heat to flow from $SA$ to the hot bath and not otherwise. 

Particularly interesting is now the analysis of the decoding and correction strokes, $W_d$ and $W_c$. 
The decoding work $W_d$ has a zeroth order contribution from the ancillas, which is  \emph{minus} the encoding work, $2p\omega$. 
If there was no noise, then the process would be entirely reversible. 
However, due to the hot bath a new contribution appears. 
This new contribution, however, appears only in the ancilla, as $W_d^S = 0$.
Moreover, this new term is always non-negative since positive temperatures imply $f \in [0,1/2]$. 
Hence, we see that the total work of the encoding/decoding process, $W_e + W_d >0$. 
It \emph{costs} work to encode and decode information when this information is scrambled by the GAD. 

The correction work $W_c$, on the other hand, is seen to be related only to changes in the system, and is  precisely minus the heat flow $Q_H^S$ between system and hot bath. 
As a consequence, the total work performed in one cycle, $W_\text{tot} = W_e + W_d + W_c$, will be
\begin{equation}
W_\text{tot} = \Omega \gamma (p-f_\Omega) + 2 \gamma \omega \bigg[ p (3-2 f_\omega) + f_\Omega (1-2p)\bigg].
\end{equation}
The second term is always non-negative, but the first term may have any sign whatsoever. 
And the  step responsible for this is the correction stroke. 
Thus, while it always costs work to encode/decode information, correcting the state may lead to either a  surplus or a deficit of work.

To linear order in $\gamma$, it follows that $W_e + Q_H + W_d + W_c + Q_C \simeq 0$. 
Referring to the first law in Eq.~(\ref{first_law}), this does not mean that the process is cyclic. 
Instead, it means that the first non-zero contribution to $\Delta U_S$ is of order $\gamma^2$:
\begin{equation}
\Delta U_S = \gamma^2 \Omega \bigg\{ f_\omega \bigg( f_\omega + 4 p - 2 p f_\omega\bigg) + 2 f_\Omega \bigg( f_\omega + p - 2 p f_\omega \bigg) - 3 p \bigg\}.
\end{equation}
Thus, even though heat and work are all of order $\gamma$, their net effect only contributes to the total change in energy with a term of order $\gamma^2$.

%
%
%
%
\section{\label{sec:shor}Shor's 9-qubit code}

\begin{figure}[!h]
\centering
\includegraphics[width=0.45\textwidth]{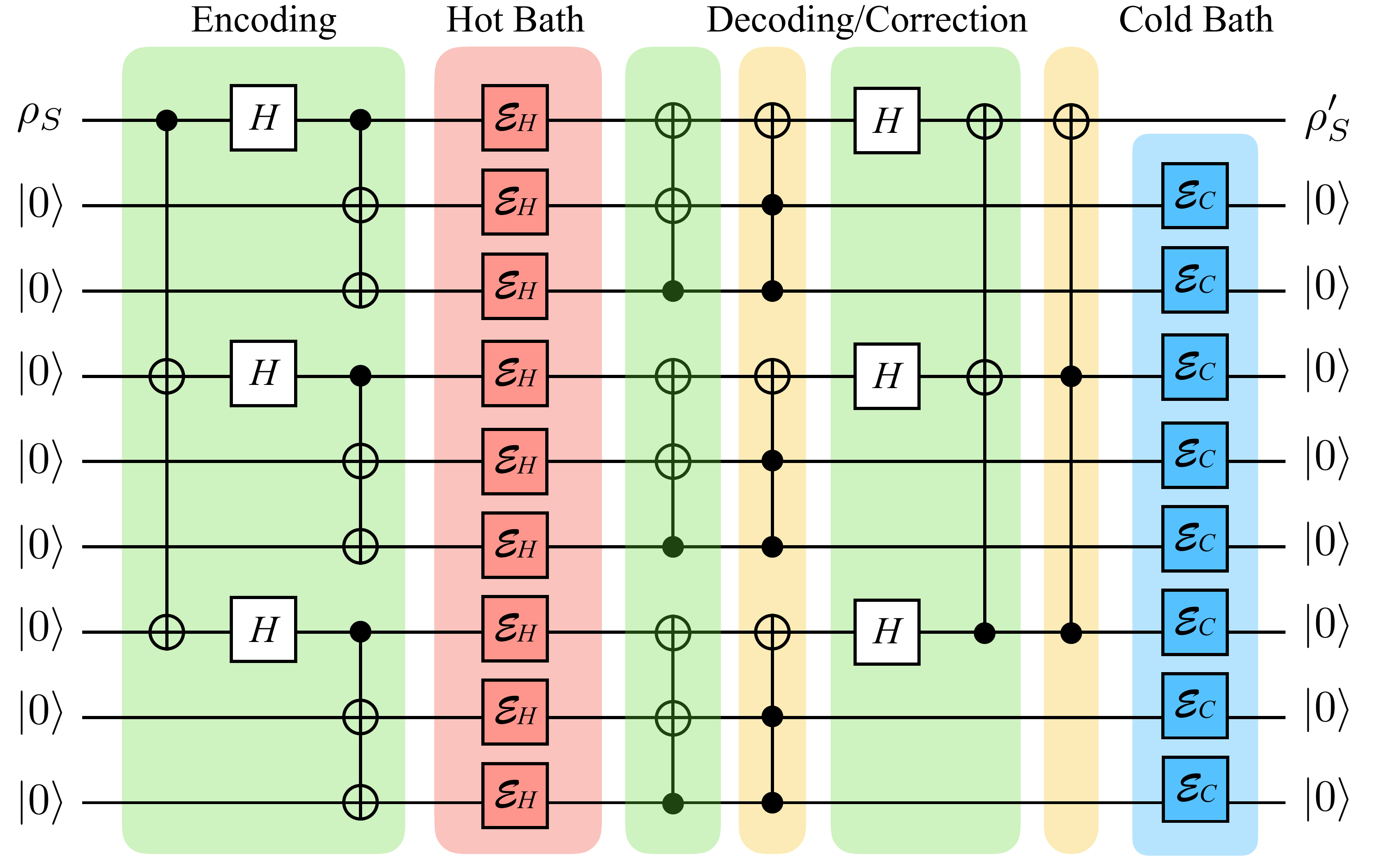}
\caption{\label{fig:shor}
Shor's -qubit code, capable of correcting both the diagonal as well as the coherent parts of $\rho_S$ against any kind of noise. 
}
\end{figure}

The 3-qubit error correcting scheme considered in the previous section is only capable of correcting diagonal states in the computational basis. 
Coherences in this basis are not correctly processed. 
A code which is capable of correcting both incoherent and coherent contributions is Shor's  famous 9-qubit code shown in Fig.~\ref{fig:shor} \cite{Shor1995} 
(see \cite{FonsecadeOliveira2017} for the implementation without syndrome measurements).
This code is quite similar in spirit to the classical code in Fig.~\ref{fig:classical}. 
The key difference, however, is that the coherent components of $\rho_S$ are also properly encoded due to the application of the Hadamard gates $H = \frac{1}{\sqrt{2}} \begin{pmatrix} 1 & 1 \\ 1 & -1 \end{pmatrix}$. 
Moreover, notice that now the decoding and correction strokes get mixed together, which we separate in Fig.~\ref{fig:shor} with different colors.

We consider a general quantum state of the system, parametrized in the form
\begin{equation}\label{parametrization_quantum_state}
\rho_S = \begin{pmatrix} p & z\sqrt{p(1-p)} \\[0.2cm] z^*\sqrt{p(1-p)} & 1-p \end{pmatrix}, 
\qquad |z| \leq 1.
\end{equation}
The state is pure when $|z| = 1$.
For simplicity, we shall also assume that $\omega = \Omega$, as the calculations become much more complex in this case. 

The work and heat in each stroke, for system and ancilla, are 
\begin{IEEEeqnarray}{rCl}
W_e^S &=& \frac{\Omega}{2}(1-2p), 		\\[0.2cm]
Q_H^S &=& - \frac{\gamma\Omega}{2} (1-2f), 		\\[0.2cm]
\label{shor_Wd_S}
W_d^S &\simeq& -\frac{\Omega}{2}(1-2p) + \frac{3\gamma \Omega }{4} (1-2p), 	\\[0.2cm]
W_c^S &\simeq& -\frac{\gamma\Omega}{4} (1+4f - 6p),	\\[0.2cm]
\end{IEEEeqnarray}
and
\begin{IEEEeqnarray}{rCl}
W_e^A &=& 4\Omega, 	\\[0.2cm]
Q_H^A &=& -4 \gamma\Omega(1-2f)		\\[0.2cm]
\label{shor_Wd_A}
W_d^A &\simeq&  - 4 \Omega + 6 \gamma \Omega(2-f)			\\[0.2cm]
W_c^A &\simeq&  \gamma\Omega(1-2f)			\\[0.2cm]
Q_C^A &\simeq& - 9 \gamma \Omega. 			
\end{IEEEeqnarray}
Several comments are worth making about these results, particularly when comparing them with the classical results in Eqs.~(\ref{classical_We_S})-(\ref{classical_QC_A}).

First and foremost, we see that all results are independent of the coherences $z$ in Eq.~(\ref{parametrization_quantum_state}). 
The reason for this is two-fold. 
First, the GAD is a thermal operation and therefore process populations and coherences independently \cite{Cwikli2015, Santos2017b}.
Secondly, the Hadamard gates in the encoding and decoding strokes (c.f. Fig~\ref{fig:shor}) acts in a way such that $z$ is not present in the reduced density matrices of a single qubit. 
Hence, since all thermodynamic quantities involve local Hamiltonians, $z$ does not appear at all in the thermodynamic aspects of the code. 

Starting with the encoding stroke, we now see that it requires work in both system and ancillas to encode information. 
Moreover, the work cost in the ancillas is entirely independent of the state of the system: for input state $\rho_S$, it will always cost the same amount $W_e^A = 4\Omega$ to encode the data in the ancillas (the work cost in the system still depends on $p$). 
A similar, but perhaps even more surprising result, is that the heat to the hot bath, for \emph{both} system and ancilla, is entirely independent of the state of the system (this state is true exactly and not to leading order in $\gamma$). 
The heat flow is simply $-\gamma \Omega(1-2f)/2$ per qubit. 
This is again a consequence of the dramatic influence of the Hadamard gates in Shor's code, which makes it so that after the encoding stroke the reduced density matrices of all qubits are simply the identity. 

The work cost of decoding is similar to the classical case [compare Eqs.~(\ref{shor_Wd_S}) and (\ref{shor_Wd_A}) with Eqs.~(\ref{classical_Wd_S}) and (\ref{classical_Wd_A})]: there is a zeroth order contribution in $\gamma$ which is simply the reverse of the encoding work (again representing the reversible part of the process). 
We also see once again that the correction work can have any sign, as in the classical case. 
And, finally, we find that the heat to the cold bath is again entirely independent of the state of the system. 

\begin{figure}[!h]
\centering
\includegraphics[width=\columnwidth]{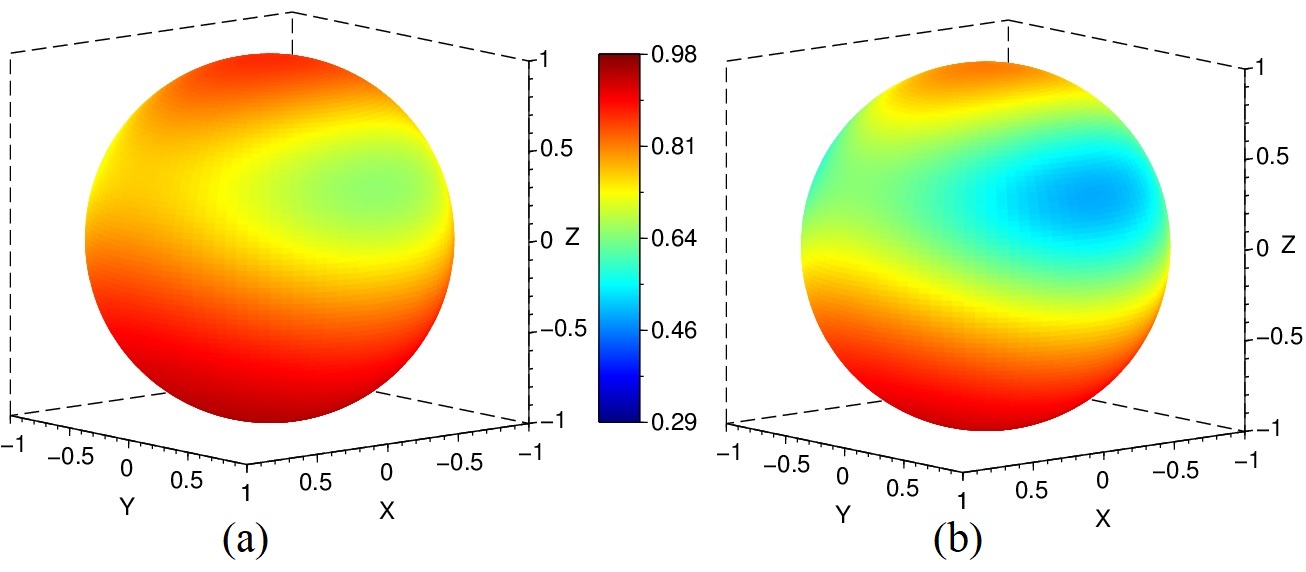}
\caption{\label{fig:shor_efic} Shor' code efficiency for input states 
on the Bloch sphere (pure states), for $f = 0.2$, $\gamma = 0.02$ (a) and $\gamma = 0.03$ (b). }
\end{figure}

On the other hand the efficiency for the Shor correcting code defined in Eq. (\ref{efficiency}) 
depends on the state, as shown in Figures \ref{fig:shor_efic}, for pure states ($|{z}|= 1$), represented as points 
of the Bloch sphere.

%
%
%
%
\section{\label{sec:disc}Discussions and Conclusions}

The framework of operator error correction (Fig.~\ref{fig:drawing}) is formally equivalent to the cyclic operation of a heat engine. 
In this paper we  aimed to explore this connection, but putting forth a thermodynamic analysis of 4-stroke codes, which parallel an Otto engine. 
We emphasize, once again, that QECC and heat engines have entirely different goals. 
In particular, for operation a QECC the work cost is only a marginal concern, as this is marginal compared to the energetics of any real experimental setup. 
That being said, the directions in which energy flows \emph{is} indeed important. Our analysis shows, for instance, that heat may very well flow \emph{from the system to the hot bath}, something which is counterintuitive. 
Indeed, this is a common misconception: neither entropy nor heat have a well defined sign. 
What does have is the \emph{entropy production}, Eq.~(\ref{Sigma}).

Another interesting aspect of this thermodynamic analysis is the interplay between the encoding and decoding work costs. 
The decoding is always the reverse of the encoding operation. 
But the effect of the noise channel in the middle of the two steps makes the process irreversible, as it scrambles information. 
As a consequence, there is always a work cost associated with the encoding+decoding steps. 

Finally, we mention an alternative perspective of the problem. 
In our formulation, the working fluid was taken to be composed of both system and ancillas, which then interacted with a hot and a cold bath. 
Alternative, one may interpret the system only as the working fluid and the ancillas as a finite sized cold bath. 
The problem with this formulation is that the system would then interact twice with this cold bath, which leads to questions related to non-Markovianity. 
The formulation as presented here is more fitting of an actual engine. 


{\it Acknowledgements.--}
The authors acknowledge fruitful correspondence with G. Guarnieri, G. Adesso,  C. Boraschi, L. Knope and L. C\'eleri. 
GTL acknowledges the S\~ao Paulo Research Foundation (FAPESP) under grant 2018/12813-0. 
GTL acknowledges Universidad ORT Uruguay, where part of this work was developed, for both the hospitality and the financial support.

\bibliography{library}

\end{document}